\documentclass[]{piparticle-final}
\usepackage{graphicx}
\usepackage{color}
\usepackage{amsmath}
\usepackage[utf8]{inputenc}

\usepackage{cite} 
\usepackage[normalem]{ulem}  
\usepackage{color} 
\usepackage{epstopdf} 

\begin{document}
\volume{8}               
\articlenumber{080004}   
\journalyear{2016}       
\editor{G. Mart\'inez Mekler}   
\reviewers{J. Mateos, Departamento de Sistemas Complejos, Instituto de F\'isica,\\ \mbox{} \hspace{3.39cm} Universidad Nacional Aut\'onoma de M\'exico, M\'exico.}  
\received{{30} October 2015}     
\accepted{4 February 2016}   
\runningauthor{G. P. Su\'arez \itshape{et al.}}  
\doi{080004}         

\title{Invited review: Fluctuation-induced transport. From the very small to the very large scales}

\author{ G. P. Su\'arez,\cite{inst1} M. Hoyuelos,\cite{inst1}\thanks{E-mail: hoyuelos@mdp.edu.ar} \hspace{0.5em}D. R. Chialvo\cite{inst2} }

\pipabstract{The study of fluctuation-induced transport is concerned with the directed motion of particles on a substrate when subjected to a fluctuating external field. Work over the last two decades provides now precise clues on how the average transport depends on three fundamental aspects: the shape of the substrate, the correlations of the fluctuations and the mass, geometry, interaction and density of the particles. These three aspects, reviewed here,  acquire additional relevance because the same notions apply to a bewildering variety of problems at very different scales, from the small nano or micro-scale, where thermal fluctuations effects dominate,  up to very large scales including ubiquitous cooperative phenomena in granular materials. }
\maketitle

\blfootnote{
\begin{theaffiliation}{99}
   \institution{inst1}Instituto de Investigaciones F\'isicas de Mar del Plata (IFIMAR - CONICET) and Departamento de F\'isica, Facultad de Ciencias Exactas y Naturales, Universidad Nacional de Mar del Plata, De\'an Funes 3350, 7600 Mar del Plata, Argentina.
      \institution{inst2} Consejo Nacional de Investigaciones Cient\'ificas y T\'ecnicas (CONICET), Godoy Cruz 2290, Buenos Aires, Argentina.
\end{theaffiliation}
}

\section{Introduction}

Much of the efforts devoted to particle transport  were triggered by the famous challenge at very small scales presented by Feynman in 1959: ``A biological system can be exceedingly small. Many of the cells are very tiny, but they are very active. (...) Consider the possibility that we too can make a thing very small, which does what we want --- that we can manufacture an object that maneuvers at that level!'' \cite{feynman}.

At the scales discussed by Feynman, our most usual notions of work, energy and transport seem to break down, including some counterintuitive observations. As discussed in these notes, these findings are not restricted to small scales since work in the last decades shows similar dynamics arising anytime there is a peculiar interplay of fluctuations, nonlinearity and correlations resulting in various classes of fluctuation-induced transport.  

To visualize the problem, consider a {\it gedankenexperiment} involving, for the sake of discussion, our desk. Elementary physics explains how all the objects at the desk stay in place and/or which forces are needed to displace them. Now, consider the imaginary case in which we progressively shrink all the objects up to a size of a few nanometers. It will be noticed that while at the natural scale objects remain steady without any energy expenditure, at the nanometers scale things move around, our ``nano cell phone'' which was quiet at the natural scale desk, moves and falls off the ``nano desk. This exercise reminds us that at the {\it Brownian domain}, energy would be required {\it even to stay quiet} since the  basic macroscopic methods of controlling energy flow no longer remain valid. This nonintuitive phenomenon in the function of molecular machines was described by Astumian as follows \cite{astumian2}: ``any microscopic machine must either work with Brownian motion or fight against it, and the former seems to be the preferable choice''. Analogous observations, with some additional caveats due to inertial forces, can be made if instead of shrinking the mass, we apply an increasingly large external fluctuating field, making now our real size desk to shake around. 

This brief review is dedicated to discuss the essence of three  elementary results of fluctuation-induced transport including the potential shape, the correlations of the fluctuations and the particle interactions and how they work, calling attention to some common lessons that can be borrowed from problems in apparently far apart scales and fields, from cellular biology to technological applications and applied physics. It should be noted that it is not our intention to cover the  extent of the field, this is neither a fair, nor historically correct, exhaustive or updated review of the relevant literature; it only encompasses some interesting results which, in our opinion, warrant further exploration. The reader will find comprehensive reviews covering specific topics, including those on Brownian motors in \cite{reimann,reimann2,astumian,parrondo,hanggi}, 
on the more general subject of molecular motors in  \cite{astumian2,Prost97,schliwa,browne,delius,chow}, on a more biological perspective of molecular springs and ratchets in \cite{mahadevan}, or on a systematic analysis of the space-time symmetries of the equations in \cite{denisov}.

The paper is organized as follows. The next section revisits pioneer works on  these types of problems, carried on a hundred years ago. Next, we discuss the three fundamental aspects of the problem, including the substrate, the correlations of the fluctuations and the particle interactions. We start by briefly introducing the different realizations of  fluctuation-induced transport as popularized two decades ago, i.e., in the so-called correlation ratchets. After that, the two elementary ways to break the symmetry are reviewed, either in the  temporal or in the spatial aspects of the system, to conclude introducing yet another way to affect transport, the correlations born out of many particle interactions. The review closes with a discussion of some  applications and new directions.

\begin{figure}
\begin{center}
\includegraphics[width=0.58\textwidth]{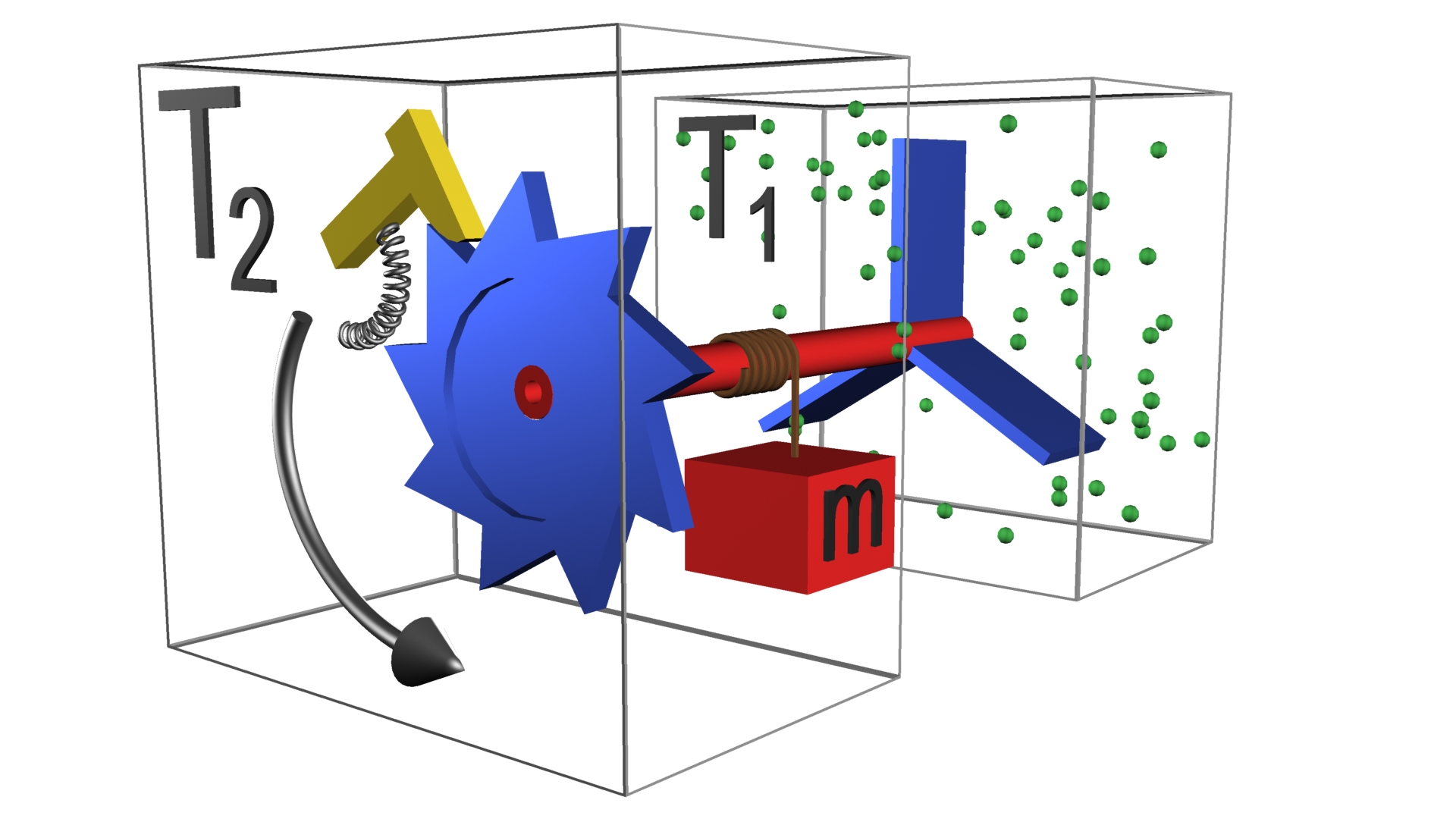}
\end{center}
\caption{ Feynman's imaginary microscopic ratchet, comprised by vanes, a pawl with a spring, two thermal baths at temperatures $T_1 > T_2$, an axle and wheel, and a load $m$. } 
\label{feynm}
\end{figure}

\section{Smoluchowski-Feynman's ratchet as a heat engine}

Feynman famous lectures \cite{lectures} include an imaginary microscopic ratchet device to illustrate the second law of thermodynamics. The basic idea belongs to Smoluchowski who discussed it during a conference talk in M\"{u}nster in 1912 (published as proceedings-article in Ref. \cite{smolu}). As seen in Fig. \ref{feynm}, it consists of a ratchet, a paw and a spring, vanes, two thermal baths at temperatures $T_1 > T_2$, an axle and wheel, and a load. The ratchet is free to rotate in one direction, but rotation in the opposite direction is prevented by the pawl. The system is assumed small so that molecules of the gas at temperature $T_1$ that collide with the vanes produce large  fluctuations in the rotation of the axle. Fluctuations are rectified by the pawl. The net effect is a continuous rotation of the axle that can be used to produce work by, for example, lifting a weight against gravity. The pawl becomes a materialization of Maxwell's demon, a small agent able to manipulate fluctuations at a microscopic level in order to violate the second law of thermodynamics, since in this case a given amount of heat is completely transformed into work. A closer inspection shows that such violation does not really take place. Feynman demonstrated that, if $T_1 = T_2$, no net rotation of the axle is produced. The reason is that the pawl has its own thermal fluctuations that, from time to time, allow a tooth of the ratchet to slip in the opposite direction. Not even demons are free from thermal fluctuations. In order for the machine to work as intended, the pawl should be colder than the vanes, $T_2 < T_1$. But in this case, there is a heat flux between thermal baths. The mechanical link between vanes and ratchet through the axle implies that the baths are not thermally isolated \cite{parrondo2}, even when the materials are perfect insulators. The system performs as a heat engine: some work is generated while some heat is transferred from a cold reservoir to a hot reservoir. 

In summary, Feynmann's ratchet ---and Brownian motors--- actually work, but without violating the laws of thermodynamics.

\begin{figure}
\begin{center}
\includegraphics[width=0.35\textwidth,clip=true]{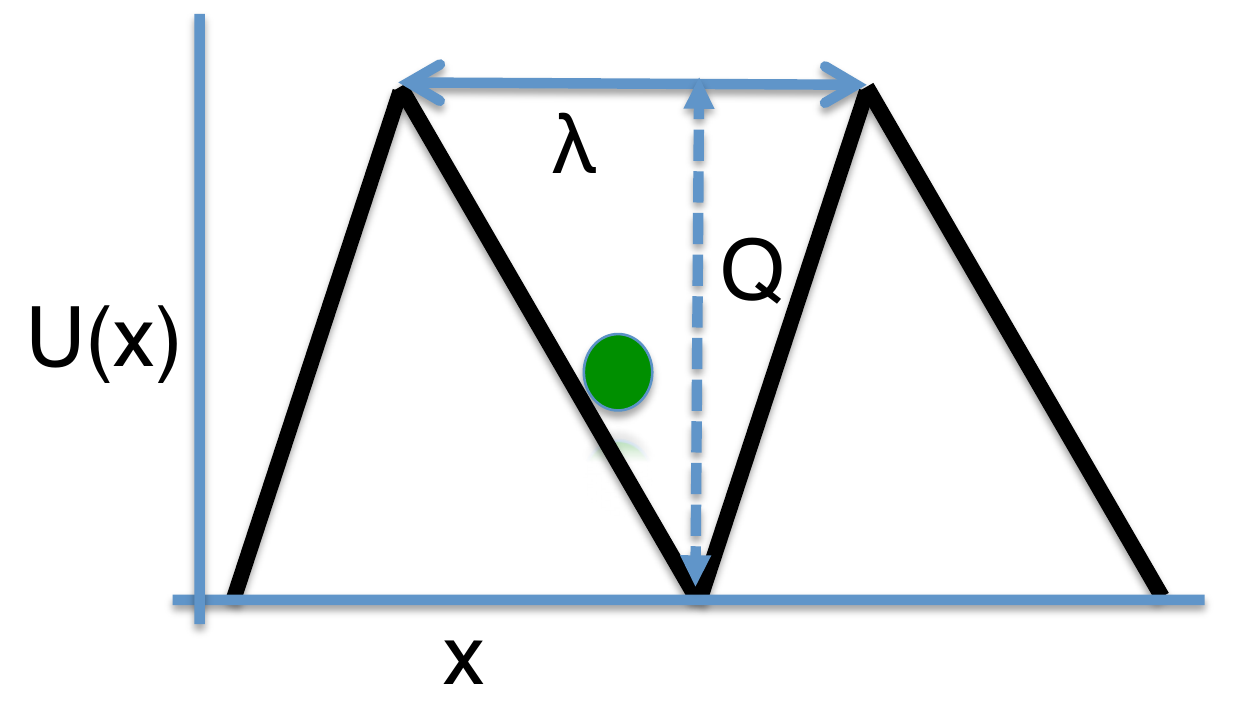}
\end{center}
\caption{Typical ratchet potential $U(x)$.} 
\label{potent}
\end{figure}

\section{Breaking the symmetry: time, space and interactions}

Feynman's deep thinking motivated an entire generation of models around the same idea. The model ratchet is a  fluctuations-driven overdamped nonlinear dynamical system described by

\begin{equation} 
\begin{split}
& \dot x = - U^\prime(x) + f(t),  \\
& U(x) = U(x+\lambda), \quad
\langle f(t)\rangle = 0,
\end{split}
\end{equation} 
where $U(x)$ is a periodic potential, such as the one illustrated in Fig. \ref{potent}, and $f(t)$ is zero-mean fluctuation of some type.  In general, the initial theoretical problem is to find the stationary current density $j = \langle \dot{x}(t)\rangle$ in the ratchet given the statistical properties of the fluctuation  $f(t)$ and the shape of $U(x)$, and to be able to determine the most efficient  conditions for the transformation of fluctuations  into a net current. 

Multiple variations and extensions of the same problem were studied in the 90s resulting on a jargon of names such as on-off ratchets \cite{Prost97}, fluctuating potential ratchets \cite{Astumian,Magnasco}, temperature ratchets \cite{Reimann96,BaoJD}, chiral ratchets \cite{Tu2004,Tu2005,vdbrk08}, and so on. In any case, three elements are always present: a particle which eventually will execute some motion and two forces, one coming from the external applied field and another given by the particular shape of the potential (i.e., the substrate where the particle resides). Thus, an isolated particle ``feels" two forces, but while such information is available to an observer, it is important to realize that the particle has no way to distinguish or separate these sources.  Thus, the break of symmetry resulting in average directed motion of a particle could come from either spatial or temporal sources. Yet, a third force needs to be considered in the cases in which the concentration of particles becomes  relevant and then particles mutual interactions are not negligible anymore, an aspect crucial to understand flow in channels. We will consider all these cases in the following sections.
 \begin{figure}[th]
\begin{center}
\includegraphics[width=0.42\textwidth,clip=true]{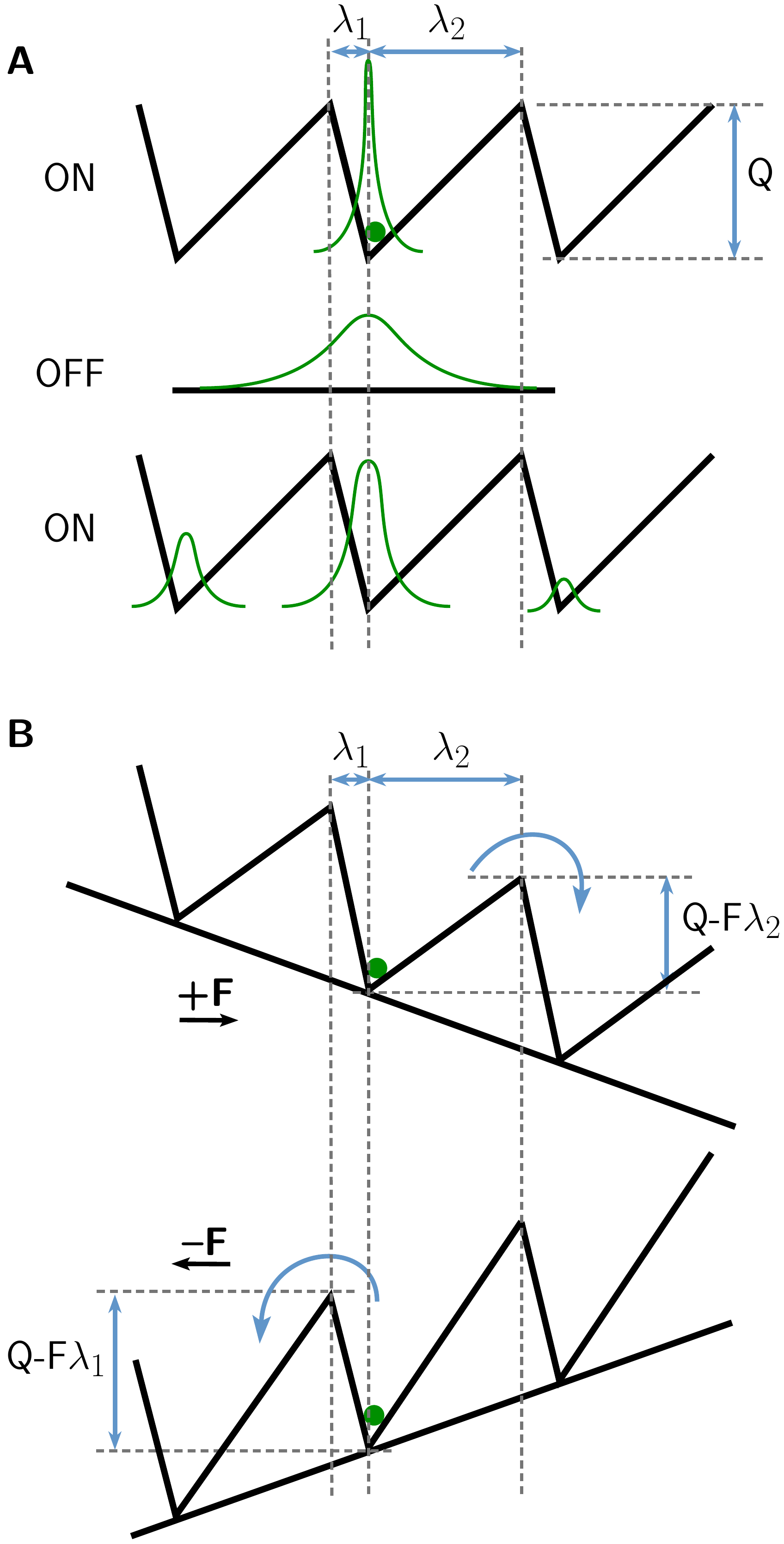}
\end{center}
\caption{Spatial asymmetry. Panel A: The so-called flashing ratchet is a type of ratchet in which an asymmetric potential  is periodically switched off and on. Particles (green circles) diffuse evenly during the off period while the asymmetric potential favors the drift in one direction, producing a net transport to the left. Panel B: In a rocked ratchet, an asymmetric  potential is tilted periodically determining a (right directed) transport of the particle trough the relative lowest $Q-F \lambda_2$ value of the right potential barrier.} 
\label{flashrock}
\end{figure}
\subsection{Asymmetries in the substrate}
  
Figure \ref{flashrock} summarizes the two basic ways in which asymmetries in space (or some other degree of freedom of the system, such as phase \cite{Millonas2}) contribute to noise-induced transport. The common situation involves an asymmetric periodic potential that breaks the spatial inversion symmetry combined with a temporal, zero mean, forcing periodicity. In Panel A, the case of  turning on and off the asymmetric potential is depicted and Panel B shows the case in which a tilting force is added.  

The first important result was due to Magnasco in \cite{Magnasco} who considered the case of the piecewise linear potential  $U(x)$ shown in Fig.\ 4A, which is exactly solvable \cite{ris} for slow fluctuation $f(t)$; it has a characteristic time much larger than the ratchet's relaxation time. The potential is periodic and extends to infinity in both directions. $\lambda$ measures the spacing of the wells, $\lambda_1$ and $\lambda_2$ the inverse steepnesses of the potential in opposite directions out of the wells, and $Q$ the well depths. The particle undergoes overdamped Brownian motion due to its coupling with a thermal bath of temperature $T$, and an external driving $F(t)$ which represents the forces. These two ingredients compose what we called the fluctuation $f(t)$. The expression for the current in the adiabatic limit, which measures the work done by the ratchet was shown to be 

\begin{align}
& J(F) ={P_2^2    \sinh(\lambda    F/2 kT)\over kT  \left(\frac{\lambda}{Q}\right)^2 P_3 - \left(\frac{\lambda}{Q}\right)P_1 P_2 \sinh \left(\frac{\lambda F}{2kT} \right)},  \\
& P_1 = \delta + {\lambda^2 -\delta^2\over      4}{F\over  Q},\ \ \ P_2=  \left[ 1 - {\delta F\over   2 Q}  \right]^2   - \left( {\lambda   F\over   2  Q} \right)^2, \nonumber \\
& P_3 =   \cosh[(Q- \delta F/2)/kT]  - \cosh(\lambda  F/2kT), \nonumber
\end{align}
where $\lambda = \lambda_1 + \lambda_2$ and  $\delta =  \lambda_1-\lambda_2$.
The  average  current,  the   quantity  of primary interest, is  given  by

\begin{equation}
j = \langle J \rangle = {1\over \tau} \int_0^\tau J(F(t))\  dt,
\end{equation}
where  $\tau$  is  the  period    of   the  driving   force $F(t)$,  which  is assumed longer than  any other   time  scale   of the   system in this  adiabatic limit. The current is maximized for a given value of the periodic forcing amplitude. Interestingly, numerical computations showed robustness in the results when the forcing is not periodic. The key feature is that it should have a long time correlation.

According to Magnasco, ``all that is needed to generate motion and forces in the Brownian domain is loss of symmetry and substantially long time correlations'' \cite{Magnasco}. Indeed, if the forcing is white noise, the system is at thermal equilibrium and  $j=0$. However, if the fluctuation auto-correlations are non-vanishing, i.e., for colored noise, the system is no longer in thermal equilibrium, and in general $j \neq 0$. Since onset of a current means breaking the ``right-left" symmetry, currents may only arise, in the case of additive noise, if the potential $U(x)$ is asymmetric with respect to its extrema. It could be argued \cite{Millonas} that the emergence of current  can be viewed  as an example of ``temporal order coming out of disorder", since the current is apparently time-irreversible, whereas stationary noise does not distinguish ``future" from the ``past"; we notice, however, that Eq.\ (1) implies relaxation and is thus time-irreversible itself.

The flashing or pulsating ratchet depicted in Panel A of Fig.\ \ref{flashrock} was introduced in \cite{bug} and re-introduced in a more general theoretical context in \cite{ajdari}. Despite the huge structural complexity of biological Brownian motors, the majority of the models are compatible with a simplified description based on the flashing ratchet. The description is in terms of only one variable $x$ that may represent, for example, the position of a molecule or the coordinate of a complex reaction with many intermediate steps. The environment, composed by some aqueous solution acts, on one hand, as a heat bath and, on the other hand, as a source or sink of ATP, ADP and P$_\text{i}$ molecules of the chemical reaction cycle that provides energy to the motor. In this simplified model, a periodic asymmetric potential is periodically turned on and off, as shown in Fig.\ \ref{flashrock}A. The situation is generalized to stochastic variations of the potential with a characteristic correlation time. As happens for the rocked ratchet, the current vanishes for zero correlation time, or fast pulsating limit (white noise). It also disappears in the slow pulsating limit, i.e., when the potential is left on or off for a diverging time. There is an optimum value of the correlation time that maximizes the current.

A recent example of an experimental realization of a rocked ratchet in a mesoscopic scale can be found in \cite{arzola}, where dielectric particles suspended in water are affected by a ratchet potential given by a periodic and asymmetric light pattern.
 
\subsection{Temporal asymmetries}
Figure 4B summarizes one type of ratchet  in which the higher order statistics  of the driving force can be responsible for the transport. Indeed, the work of Millonas \cite{Millonas1,Millonas2,Millonas3} and others \cite{Mahato} showed that directed motion can be induced with an unbiased driving force, deterministic or stochastic, as long as it has asymmetric correlations: non zero odd correlation of order higher than one \cite{Millonas}.
 \begin{figure}[th]
\begin{center}
\includegraphics[width=0.35\textwidth]{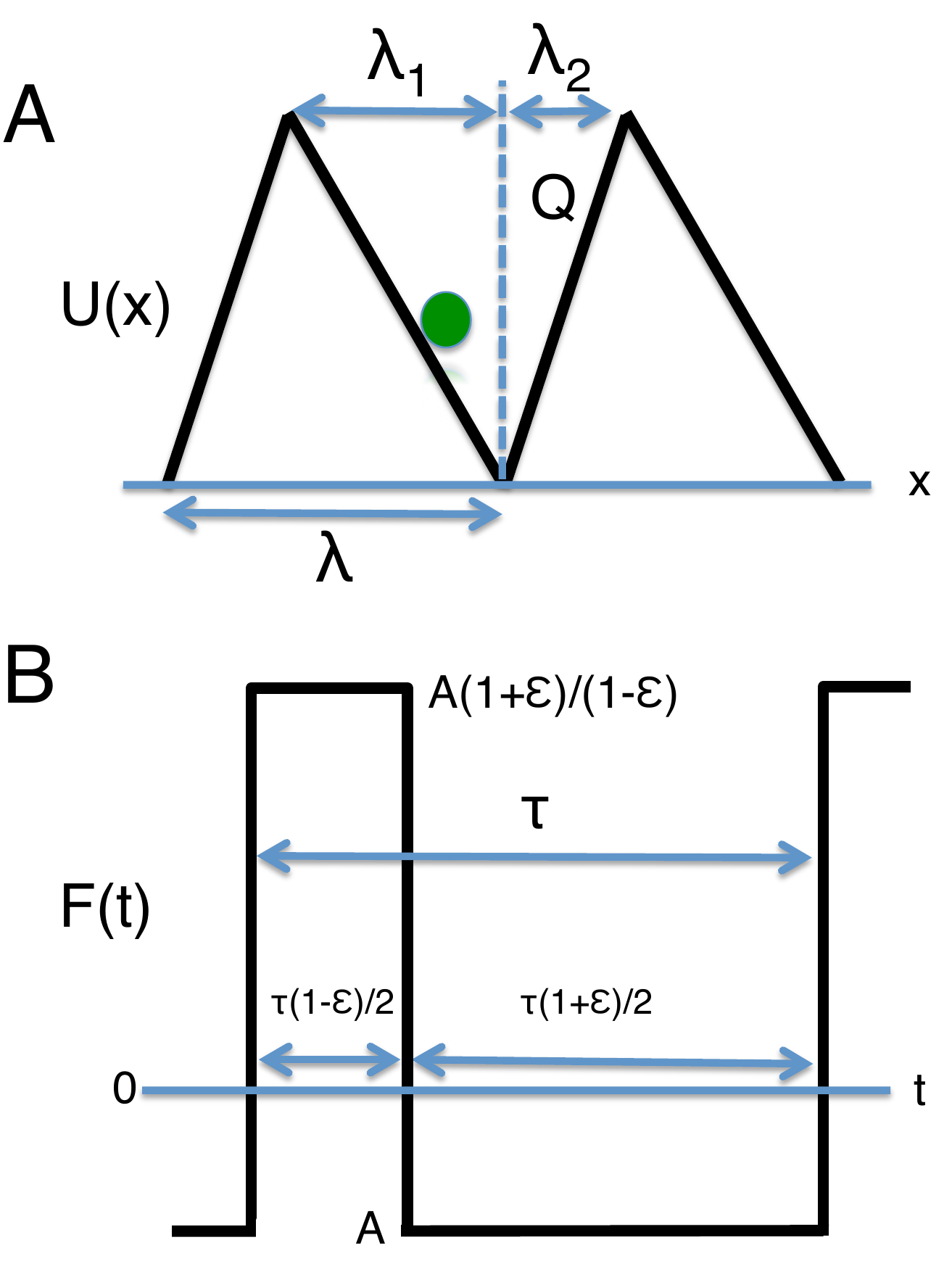}
\end{center}
\caption{ Panel A: The simplest piecewise ratchet potential, where the spatial degree of asymmetry is given by the parameter $\delta=\lambda_1- \lambda_2$. Panel B: Fluctuation's temporal asymmetry. The driving force $F(t)$ preserves the zero mean $\langle F(t)\rangle=0$.  The temporal asymmetry is given by the parameter $\epsilon$.} 
\label{ratasym}
\end{figure}

The case analyzed in the seminal work of Magnasco \cite{Magnasco}  only considered $F(t)$ symmetric  in time  $F(t)=F(n\tau-t)$.   Instead, the work of Millonas \cite{Millonas1} considered  the same setting but studying a more general case in which the driving  force still is non biased  {\it zero mean}, $\langle F(t)\rangle \ =0$, but  which is asymmetric in time,

\begin{equation}
F(t) = 
\left\{ \begin{array}{ll}
\left({1+\epsilon\over 1-\epsilon}\right)A 
&  0 \leq  t< {\tau (1 -\epsilon)\over 2}, \mod \tau \\
 - A &  {\tau (1-\epsilon)\over 2} < t \leq \tau, \mod \tau
\end{array}
\right.
  \end{equation}
as shown in Fig. 4B. In this case, the  time averaged current  can be  easily calculated,
\begin{equation}\begin{split}
\langle J\rangle =&   {1\over 2} (1+\epsilon) J(-A) \\
  &+ \frac{1}{2}(1-\epsilon)J((1+\epsilon) A/(1-\epsilon))    
\end{split}
\end{equation} 
\begin{figure}[th]
\begin{center}
\includegraphics[width=0.35\textwidth]{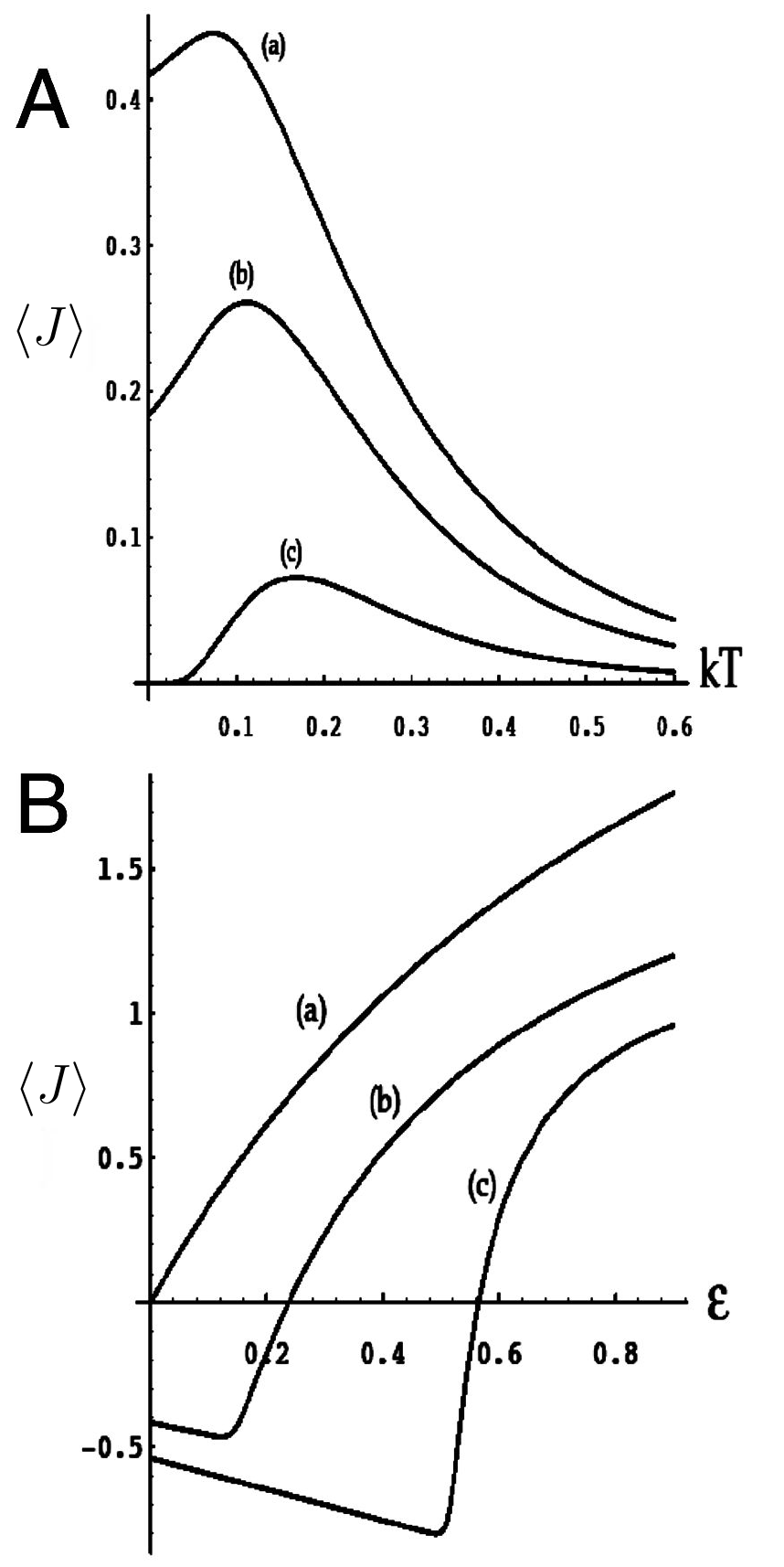}
\end{center}
\caption{Temporally asymmetric fluctuations  with mean zero can optimize or reverse the current in the ratchet of Fig. 4 (with $Q=1,\lambda =1$). Panel A corresponds to the case of a symmetric potential (i.e., $\delta=0$) which shows a peaked function of the net current $\langle J \rangle$ as a function of temperature $kT$ for $\epsilon =1$ and three values of the driving amplitude $A=1, A = 0.8, A = 0.5$ labeled (a), (b) and (c), respectively.
Panel B shows $\langle J \rangle$ versus the temporal asymmetry parameter $\epsilon$  for three asymmetries in the shapes of the potential. Curve (a) is for a symmetric potential, (i.e., $\delta =0$) and those labeled (b) and (c) for two cases of asymmetric potentials $\delta = - 0.3$  and $\delta = - 0.7$, respectively (with $kT = 0.01$, and $A = 2.1$) }
\end{figure}

Solving for different values of parameters, it was shown that the current is a peaked function both of $kT$ (see Fig. 5A) and of the amplitude $A$ of the driving.  As expected, the driving, the potential, and the thermal noise in fact play cooperative roles. For low temperatures,  any transport depends  on very large $A$ values, while  for large noise the features of the potential and of the driving are washed out. 

The most striking results are concerning the competition between the temporal asymmetry and the spatial asymmetry, as pictured in Fig. 5B, resulting on the switching of the direction of the current as  the asymmetry factor $\epsilon$ is varied.  This reversal represents the competition of the spatial asymmetry, which dominates for small $\epsilon$ an the temporal asymmetry, which dominates for large $\epsilon$.

Temporal asymmetry and spatial asymmetry relate to the problem of nonequilibrium transport in precisely the same way. In both cases, a net effect arises  due to an interplay between the strength of a fluctuation, the time it acts, and the underlying dynamics. In the case of a spatial asymmetry, a fluctuation to the right with a given strength which lasts a given time will tend to take the system over the right-hand barrier while the same fluctuation with sign reversed does not lift it over the left-hand barrier. In the case of temporal asymmetry, the  probabilities of the fluctuations to the right or to left are different, so the net effect arises in the absence of spatial asymmetries. What both of them show is that even a subtle asymmetry in the {\it shape} of the potential or in the {\it shape} of the spectral properties of the noise will give rise to an effect even when the net force due to each vanishes.

The time asymmetry of the mean zero fluctuations discussed above can be cast in several different ways. Dichotomic noise (a type of ``Kubo-Anderson" process)  was used to demonstrate phase transport in a pair of Josephson junctions \cite{Millonas2}. There are also types of continuous noise exhibiting similar asymmetry, including shot noise (common in quantum electronics) which are of this type. Mean zero shot noise, which is temporally asymmetric, can be produced if the frequency and amplitude distribution are slightly different for positive and negative fundamental pulses.  Another trivial example of temporally asymmetric driving force is a simple bi-harmonic signal which constitutes a curiosity since it results from adding two (zero mean symmetric) periodic process of harmonic frequencies.

\subsection{Particle interactions}

The previous discussions were limited to the cases in which an isolated or a few particles were present in the potential. As the concentration is increased, interaction among particles becomes relevant, and it can be the cause of a reduction, and even of a reversal \cite{kostur,savelev,ai,derenyi} of the current. We present two examples.

\begin{figure*}
\begin{center}
\includegraphics[width=0.9\textwidth]{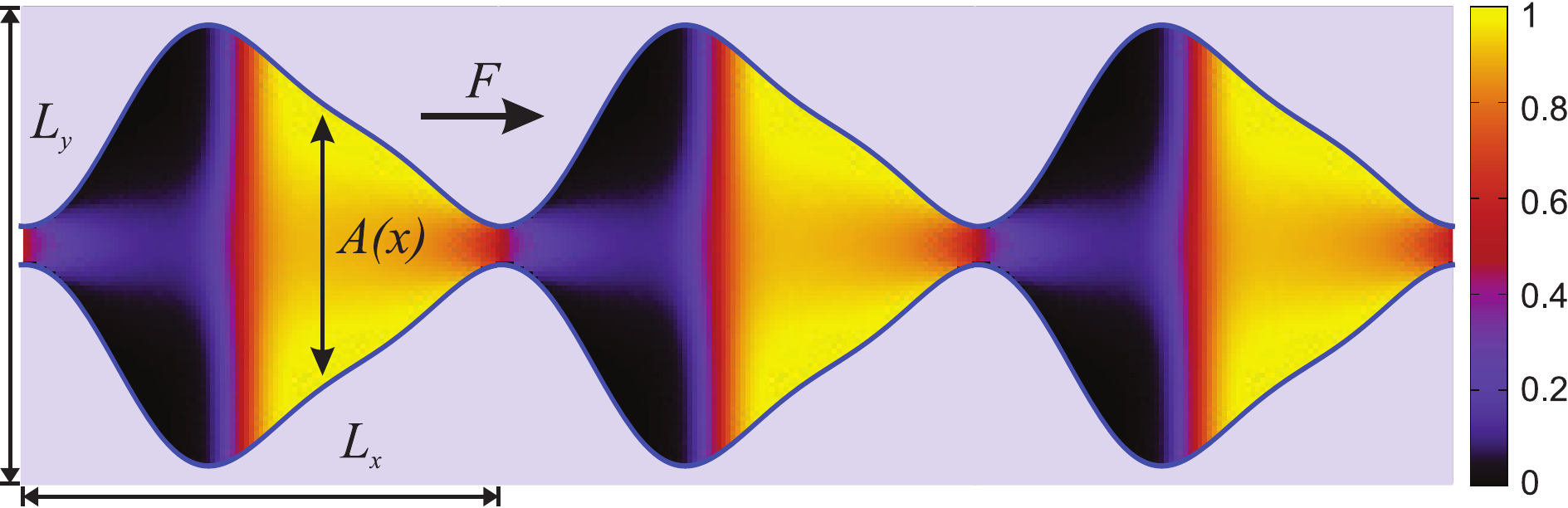}
\end{center}
\caption{Diffusion in a periodic channel with asymmetric cavities when a force to the right is applied; $A(x)$ is the witdth of the channel. Parameters: total average concentration $c=0.5$; force $\beta\,a\,F = 0.5$, where $a$ is the lattice spacing; $L_x = L_y = 100 \, a$.} \label{cavity}
\end{figure*}

\subsubsection{Vortex current in a 2D array of Josephson junctions}

Current reversal has been experimentally observed in a two dimensional array of Josephson junctions \cite{shalom}. It was numerically analyzed in \cite{marconi}. A ratchet potential for vortices is generated by modulating the gap between superconducting islands. The density of vortices is controlled by an external magnetic field. There is a repulsive vortex-vortex interaction. The results show a preferred direction of the vortex motion, parallel to the ratchet modulation, when an alternating force is applied. But as the vortex concentration is increased, this direction is reversed for appropriate values of the periodic forcing intensity. (Vortex current reversal is also observed for a fixed value of the concentration when the periodic forcing, or AC current amplitude, is varied, see Fig. 4 in \cite{shalom} or Fig. 2 in \cite{marconi}).

The vortex current reversal produced by the increase of concentration is a consequence of the following symmetry \cite{shalom}. Let us consider that the external magnetic field is such that positive vortices are produced. For small concentration (frustration parameter between 0 and 1/2), we have a small discrete number of positive vortices. For large concentration (frustration between 1/2 and 1), we can consider that there is a background of positive vortices in which some negative vortices move. But the movement of this negative vortices is in the opposite direction. For them, the ratchet potential is inverted so the rectification effect of the ratchet is inverted too.

\subsubsection{Particle diffusion in a channel with asymmetric cavities}

The same effect is observed in a different context. In the next paragraphs, we refer to the hard core interaction between particles that diffuse in a channel with a transverse section $A(x)$ that has a ratchet shape, see Fig.\ \ref{cavity}. An external periodic forcing is applied in the direction of the channel. There is a particle-hole symmetry. But before going into the interaction effects, let us consider the low concentration regime, where interactions can be neglected. Several interesting experiments have been performed with particles suspended in a liquid and contained in a channel qualitatively as the one shown in Fig. \ref{cavity}. There are basically two ways to apply the periodic external forcing. In one case, a periodic variation of the pressure is used: particles are drifted back and forth by the movement of the liquid; see \cite{kettner,matthias} and the critical report \cite{mathwig} (cavities of order 5 $\mu$m). In the other case, the liquid remains still and the force is directly applied on the particles by an external field as, for example, an electric field on charged particles \cite{marquet} (cavities of order 50 $\mu$m). 

Such a system has been proposed for separation of particles of different size \cite{reguera}. The idea is based on the difference between rectification effects for different size particles. When a periodic ---unbiased--- forcing is applied, particles move in the forward direction because of the ratchet; but, in general, larger particles move faster than smaller ones. Now we apply a bias, a constant force in the backward direction that reduces the velocity of the larger particles and \textit{reverses} the velocity of the smaller ones. Then we have that larger particles end up in one extreme of the channel and smaller particles in the opposite one, with an estimated purity of 99.997 \% according to the authors of \cite{reguera}.

The Fick-Jacobs equation \cite{jacobs} gives an appropriate description of the particle density in the channel as long as its dependence on the transverse direction, $y$, can be neglected, i.e., $n(x,y,t) \simeq n(x,t)$. Let us consider the transverse integral of the concentration: $\rho(x,t) = \int dy\, n(x,y,t) \simeq A(x) n(x,t)$ (the two-dimensional channel can be easily extended to a three-dimensional tube). If $D$ is the diffusion coefficient and $F(t)$ is the total applied force, the Fick-Jacobs equation is

\begin{equation}
\frac{\partial \rho(x,t)}{\partial t} = \frac{\partial}{\partial x}\left[ D \,\left( \frac{\partial \rho}{\partial x} + \beta \frac{\partial H}{\partial x}\, \rho\right) \right],
\end{equation}
where $\frac{\partial H}{\partial x} = -F(t) - \beta^{-1} \frac{d}{dx}\ln A(x)$ and $\beta^{-1} = k T$. The expression $\beta^{-1}\ln A(x)$ is called entropic potential due to the similarity with the thermodynamic relation among energy, free energy, temperature and entropy: $H = U - TS$, with $\frac{\partial U}{\partial x}= -F(t)$. In a first approximation, the diffusion coefficient is constant; a further refinement considers a dependence on $A'(x)$, see \cite{zwanzig}. 

\begin{figure}
\begin{center}
\includegraphics[width=0.5\textwidth]{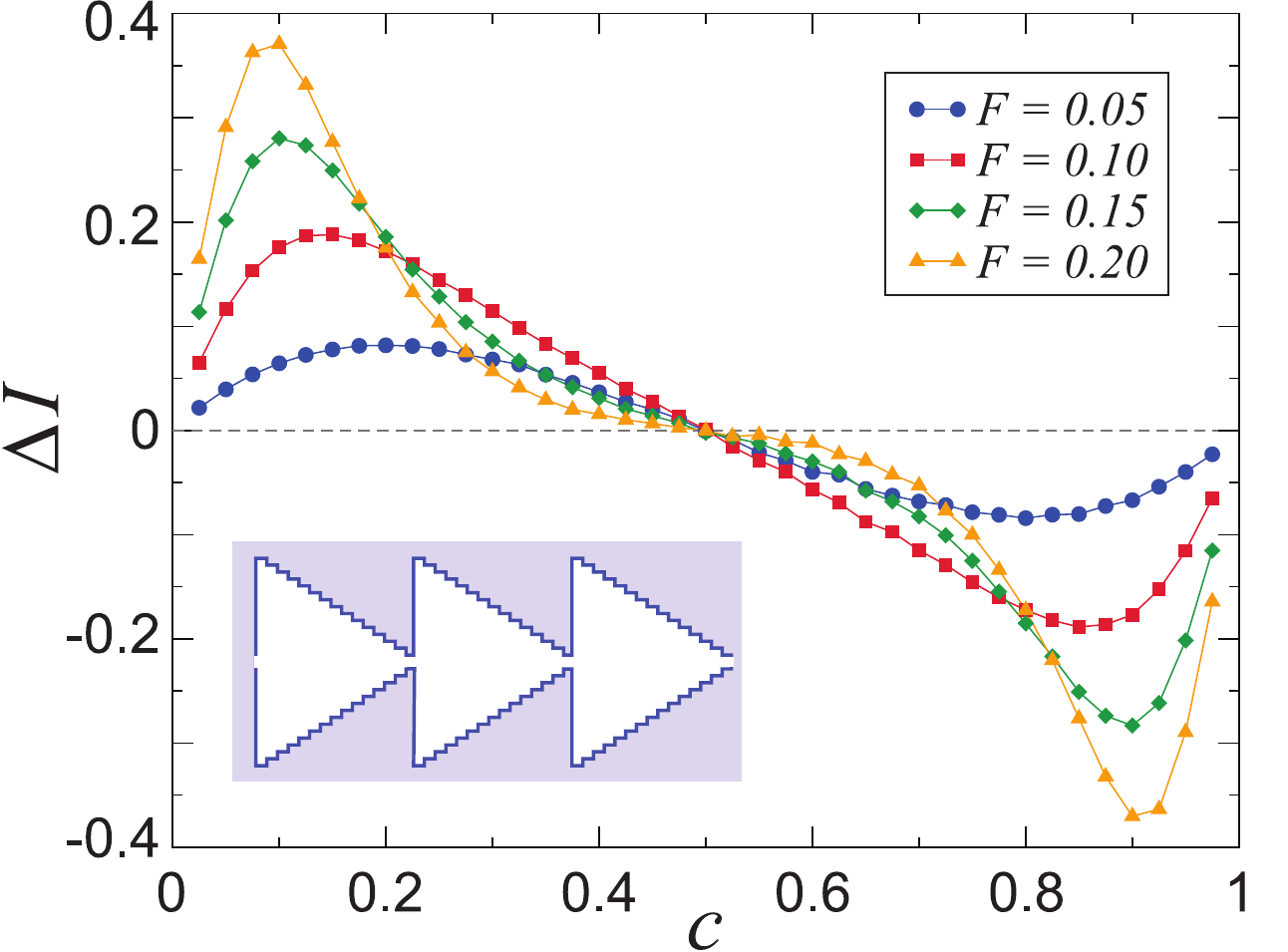}
\end{center}
\caption{Rectified particle current $\Delta I$ against average particle concentration $c$ for different force amplitudes. Units of $\Delta I$: $D/a^2$, where $D$ is the diffusion coefficient and $a$ is the lattice spacing; units of $F$: $(\beta a)^{-1}$. Lower left inset: scheme of the channel composed by an array of triangular cavities.} \label{currdiff}
\end{figure}

Now, let us consider the hard core interaction between particles, of the same size, diffusing in a lattice. A jump of a particle to the right is equivalent to a jump of a hole to the left. A concentration $c$ of particles subjected to a force $F$ is equivalent to a concentration $1-c$ of holes subjected to a force $-F$. This symmetry is the cause of the shape of Fig. \ref{currdiff}, where Monte Carlo results of the rectified current $\Delta I$ against the average concentration $c$ for different values of the forcing amplitude is plotted \cite{suarez1}. A square wave in the limit of low frequency was used for the applied force; in this limit, the rectified current is equal to the difference between the current for the force in the positive phase and the current for the force in the negative phase. The particle-hole symmetry is evident in the figure: changing $c\rightarrow 1-c$ and $\Delta I\rightarrow -\Delta I$ (a consequence of the change $F\rightarrow -F$), we recover the same curves. Le us note the current reversal for large concentration. It is the same effect that was mentioned in the previous section for vortex current in 2D arrays of Josephson junctions.
\begin{figure}
\begin{center}
\includegraphics[width=0.5\textwidth]{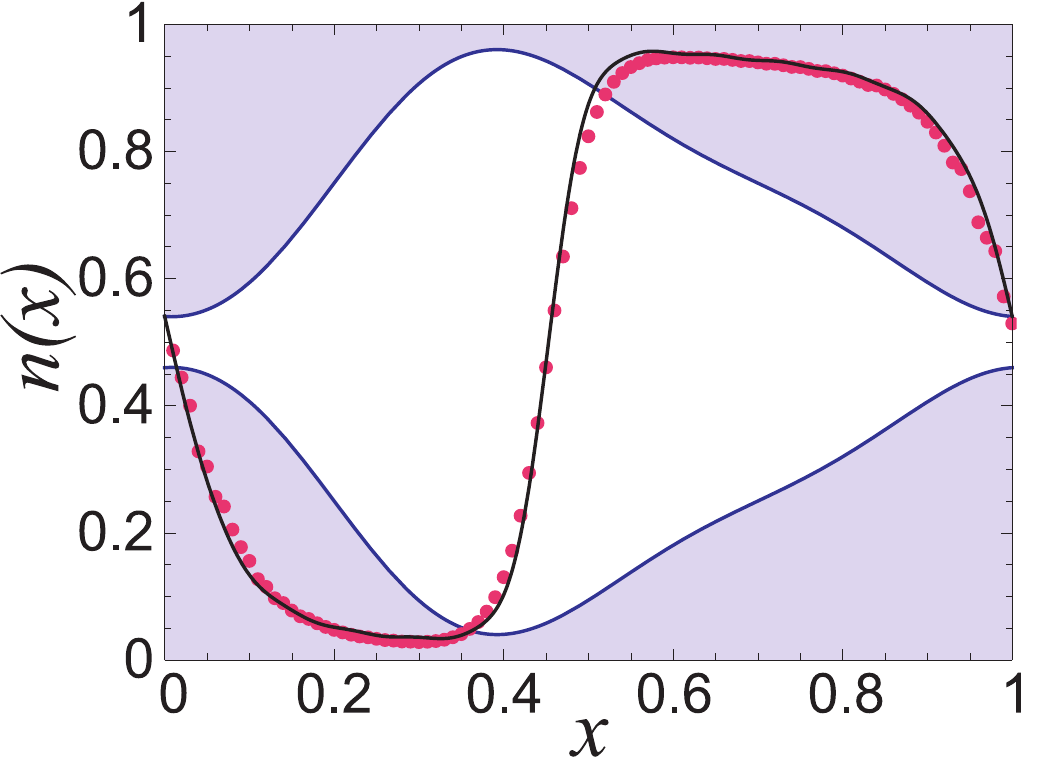}
\end{center}
\caption{Particle concentration against longitudinal position $x$ for one cavity of the channel depicted in Fig. \ref{cavity}, in a stationary state (cavity length normalized to 1). Dots correspond to Monte Carlo simulations. The curve is obtained from numerical integration of \eqref{fj}. Concetration $c=0.5$, more details in \cite{suarez2}.} \label{fjnl}
\end{figure}
A description based on the Fick-Jacobs equation is also possible for particles with hard-core \cite{suarez2}. Its derivation starts from the non-linear Fokker-Planck equation for fermions \cite{frank}, where Pauli exclusion principle plays the role of the hard core interaction. Following the same steps used for derivation of the linear Fick-Jacobs equation \cite{zwanzig}, we can arrive at the following non linear version: 

\begin{equation}
\frac{\partial n(x,t)}{\partial t} = \frac{1}{A} \frac{\partial}{\partial x} D A \left( \frac{\partial n}{\partial x} - \beta F n\,(1-n) \right).
\label{fj}
\end{equation}
The non linear term, $n\,(1-n)$, is the responsible of the interaction. Fig. \ref{fjnl} shows numerical integration of \eqref{fj} in good agreement with Monte Carlo simulation results. 

\section{Engineers knew it...}

The principles discussed above ruling the correlation ratchets have, in a macro scale, important technological applications. Of course, some of them were applied even before a detailed statistical understanding was available. Vibratory conveyors, or vibratory bowl feeders, are regularly used in many branches of industry such as food processing, synthetic materials or small-parts assembly mechanics, to mention just a few \cite{kruelle}. The conveying speed of these devices was theoretically and experimentally studied, for example, in \cite{sloot} and references cited therein.

There are many parameters involved in the operation of a vibratory conveyor: amplitude, frequency and mode of the vibrations, inclination angle and friction coefficient are only some of them. A classification in terms of the vibratory modes is as follows: sliding (linear horizontal vibration), ratcheting (linear vertical vibration) or throwing (circular, elliptical or linear tilted vibration).

\begin{figure}
\begin{center}
\includegraphics[width=0.5\textwidth]{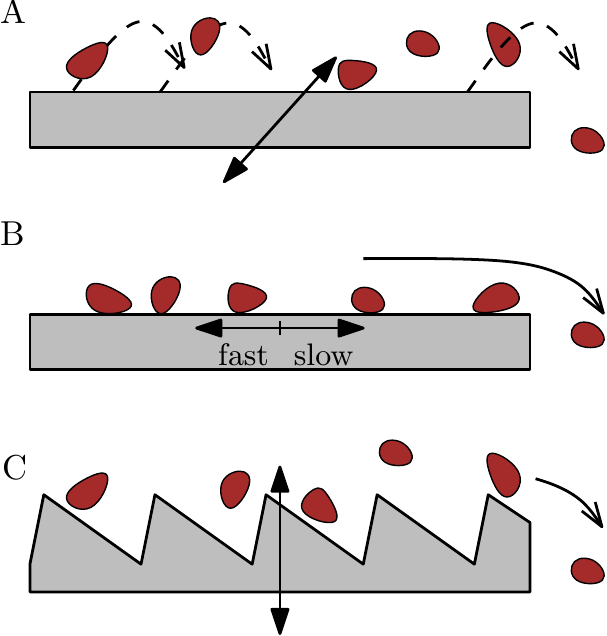}
\end{center}
\caption{Three types of vibratory conveyors which share some of the principles of small scale ratchets. Panel A: throwing conveyor with  linear tilted vibration. Panel B: sliding conveyor; transport is induced by asymmetric horizontal oscillations with zero mean, of the kind shown in Fig.\ \ref{ratasym}B. Panel C: ratchet conveyor with vertical oscillations; similar to the flashing ratchet of Fig.\ \ref{flashrock}A.} 
\label{conveyors}
\end{figure}

For throw conveyors, the material being transported loses contact with the through during part of the cycle, see Fig.\ \ref{conveyors}A. Appropriate for granular materials o small objects, particles are forced to perform repeated short flights with a preferred direction, combined with rest and slide phases.

The other two types: sliding and ratcheting involve temporal and spatial asymmetries, respectively.

The sliding type of vibratory conveyors allows transport over a deck that vibrates back and forth with asymmetric motion (see Fig.\ 5A) in the horizontal direction. The particle or object moves relative to the deck due to alternate stick and slip steps driven by the asymmetric oscillations, as shown in Fig.\ \ref{conveyors}B.

The ratchet conveyor achieves transport of granular material using vertical vibrations \cite{derenyi,farkas}. Directed motion is caused by the broken space symmetry of the deck's surface, given by a sawtooth-shaped profile, see Fig.\ \ref{conveyors}C. The ratchet conveyor shares qualitative features with the flash or pulsating ratchet depicted in Fig.\ \ref{flashrock}A. One difference is that it includes a ballistic flight phase.

\section{Conclusions}

Spatial or temporal asymmetries, or both, are able to generate directed motion in the presence of fluctuations. In addition to a thermal bath, fluctuations with large correlation time, compared to the characteristic relaxation time of the system, should be included. During the last decades, simple models based on these ideas provided a deeper understanding of the complex biological machinery at the nano scale. This success stimulated the study of ratchets in a wide variety of contexts, and in larger scales. Interactions among transported particles are relevant for high concentration; most noticeable, they may produce an inversion of the purported motion direction.

Vibrations inducing directed motion are used in industry for the transport of small ---macroscopic--- objects since, at least, around 1950. Vibratory conveyors applied the qualitative features of ratchets, with space or time asymmetries, before a detailed theoretical understanding was available.
Half a century ago, Feynman called attention to the fact that in his view, ``there's plenty of room at the bottom'' \cite{feynman}. We can safely conclude that, even today, there is plenty of room at the top as well.


\begin{acknowledgements}
This work was partially supported by Consejo Nacional de Investigaciones Cient\'ificas y T\'ecnicas (CONICET, Argentina)
\end{acknowledgements}

\end{document}